\documentstyle[aps,twocolumn,floats]{revtex}
\input psfig.sty
\begin{document}
\draft
\title{Optical absorption of spin ladders}
\author{Christoph Jurecka$^\dagger$ and Wolfram Brenig$^{\dagger,\star}$}

\address{$^\dagger$Institut f\"ur Theoretische Physik,
Technische Universit\"at Braunschweig, 38106 Braunschweig, Germany}

\address{$^\star$Institute for Theoretical Physics,
University of California at Santa Barbara, CA 93106-4030, USA}
\date{\today}
\maketitle
\begin{abstract}
We present a theory of phonon-assisted optical two-magnon absorption in
two-leg spin-ladders. Based on the strong intra-rung-coupling limit we
show that collective excitations of total spin $S=0,1$ and $2$ exist
outside of the two-magnon continuum. It is demonstrated that the singlet
collective state has a clear signature in the optical spectrum.
\end{abstract}

\pacs{
71.27.+a, 
75.10.Jm, 
75.50.Ee, 
75.30.Et, 
78.30.Hv  
\\ ITP preprint: NSF-ITP-99121}

Recent design of materials with localized spin-1/2 moments arranged in
ladder, as well as frustrated and dimerized chain geometries has enormously
intensified research on low-dimensional quantum magnetism
\cite{Dagotto99}. Many of the novel materials are likely candidates
for a spin-liquid ground state and display a gapped spin-excitation
spectrum. On the two-leg ladder the spin-gap is a phenomenon well
established by various theoretical approaches, i.e.  exact diagonalization
\cite{Barnes93,Barnes94}, strong coupling expansion
\cite{Reigrotzki94,Oitmaa96}, density-matrix renormalization group
\cite{White94}, and bosonization \cite{Shelton96}.  Optical spectroscopy, and
in particular Raman scattering, has proven to be a valuable tool to
investigate the spin excitations of the new quantum magnets
\cite{Lemmens99}. Apart from Raman scattering, optical absorption has has
been studied, however, only on the {\em single}-chain cuprate Sr$_2$CuO$_3$
\cite{Suzuura96,Eder96} where it has provided direct evidence for the gapless
two-spinon spectrum of the spin-1/2 Heisenberg chain \cite{Bougourzi98}. In
Sr$_2$CuO$_3$ absorption is due to phonon assisted two-magnon emission
(PME) \cite{Lorenzana95a,Lorenzana95b}.  While several of the new
spin-ladder materials, eg. SrCu$_2$O$_3$ and LaCuO$_{2.5}$, may allow for
PME as well, corresponding studies are lacking. In this work we propose a
simple theory of PME for spin-ladders. We show that {\em bound states} in
the spin-gap which have no analogy in the Heisenberg chain have a
substantial effect on the optical absorption spectrum.

We start our discussion with the two-leg spin-ladder Hamiltonian
\begin{equation}
\label{w0}
H=\sum_{l,\alpha} [S^\alpha_{1 l} S^\alpha_{2 l}
+\lambda (S^\alpha_{1 l} S^\alpha_{1 l+1}+
S^\alpha_{2 l} S^\alpha_{2 l+1})]
\end{equation}
where $S^\alpha_{\mu l}$ with $\alpha=x,y,z$ is a spin-1/2 operator on
site $l$ of leg $\mu$ and H is measured in units of $J_\perp$ with
$\lambda=J_\parallel/J_\perp$. In this paper we focus on the limit of
large intra-rung coupling $\lambda\rightarrow 0$ in which a reformulation
of (\ref{w0}) in terms of the rung-spin $S^\alpha_l=S^\alpha_{1 l}+
S^\alpha_{2 l}$ and the operator $T^\alpha_l=S^\alpha_{1 l}-S^\alpha_{2 l}$
is appropriate
\begin{eqnarray}
H=\sum_{l}[&&-\frac{3}{4}+\frac{1}{2} S_l(S_l+1)]
+\frac{\lambda}{2}\sum_{l,\alpha} T^\alpha_lT^\alpha_{l+1}
\nonumber\\ \label{w1}
&&+\frac{\lambda}{2}\sum_{l,\alpha} S^\alpha_lS^\alpha_{l+1} =
H_1 + H_2 + H_3
\end{eqnarray}
where $S_l=0(1)$ for a rung singlet(triplet). The rung-spin eigenbasis is
given by
$|s\rangle = (
|\uparrow\downarrow\rangle-|\downarrow\uparrow\rangle)/\sqrt{2}$,
$|t_x\rangle = -(
|\uparrow\uparrow\rangle-|\downarrow\downarrow\rangle)/\sqrt{2}$,
$|t_y\rangle = i(
|\uparrow\uparrow\rangle+|\downarrow\downarrow\rangle)/\sqrt{2}$, and
$|t_z\rangle = (
|\uparrow\downarrow\rangle+|\downarrow\uparrow\rangle)/\sqrt{2}$
where the first(second) entry in the kets refers to a site on leg '$1(2)$'
of the ladder.  As usual $S^\alpha|s\rangle=0$ and $S^\alpha|t_\beta\rangle
=i\varepsilon_{\alpha\beta\gamma}|t_\gamma\rangle$ with the Levi-Civita
symbol $\varepsilon_{\alpha\beta\gamma}$. The action of $T^\alpha$ on the
rung basis is given by $T^\alpha |s\rangle = |t_\alpha\rangle$ and
$T^\alpha |t_\beta\rangle = \delta_{\alpha\beta}|s\rangle$.

For vanishing inter-rung coupling the ground state of (\ref{w1}) is a pure
rung-singlet product-state $|\rangle = \otimes_{l} |s_l\rangle$. The excited
states are products of singlets and triplets $|\{t_{m\alpha}\}\rangle =
\otimes_{\{l\}} |s_l\rangle \;\otimes_{\{m\alpha\}} |t_{m\alpha}\rangle$ with
an excitation energy given by the number of triplets $N[\{t_{m\alpha}\}]$. At
finite $\lambda$ the action of the ladder-Hamiltonian on these product states
can be read off easily from (\ref{w1}): (i) $H_2$ creates(destroys) pairs of
nearest-neighbor (NN) triplets of equal $\alpha$-index, (ii) given a pair of
NN sites in a relative state of one singlet and one triplet $H_2$ generates
NN hopping of the triplet via exchange of the singlet and triplet, and
finally (iii) $H_3$ induces NN interactions between triplets. Contributions
(ii) and (iii) define a three-flavored hard-core Bose system with NN hopping
as well as NN interactions. These contributions renormalize the spectrum to
$O(\lambda)$. Process (i) does not conserve the triplet number and changes
the spectrum only to $O(\lambda^2)$ at $\lambda\ll 1$. The essential physics
of the bound states and optical absorption is independent of
$O(\lambda^2)$-terms.  They will be neglected hereafter.  Based on this
simplification it is convenient to split the Hamiltonian into a {\em bare}
'kinetic' part $H_1+H_2$ and a two-particle interaction $H_3$. The {\em bare}
one-triplet eigenstates $|k\alpha\rangle$ of momentum $k$ are
\begin{eqnarray}
\label{w5}
|k\alpha\rangle & = & \frac{1}{\sqrt{N}}\sum_l e^{ikl}
|t_{l\alpha}\rangle
\makebox[2cm]{}
\\
\label{w5b}
\varepsilon_k & = & \langle k\alpha|H_1+H_2|k\alpha\rangle -
\langle |H|\rangle = 1+ \lambda \cos(k)
\end{eqnarray}
where $\langle k'\alpha|k\beta\rangle=\delta_{\alpha\beta}
\delta_{kk'}$. This agrees with seminal work on spin-ladders
\cite{Barnes93,Barnes94,Reigrotzki94}. Hereafter the ground state energy
$\langle |H|\rangle$ is set to zero. The {\em bare} two-triplet eigenstates
in the singlet sector must incorporate the symmetry
$|t_{l\alpha}t_{m\beta}\rangle=|t_{m\beta}t_{l\alpha}\rangle$ as well as
the hardcore constraint $|t_{l\alpha}t_{l\beta}\rangle=0$
\begin{eqnarray}
|kq,S\rangle = && \frac{1}{\sqrt{N(N-1)}} \sum_{l,m} [
e^{ik(l+m)/2}
\nonumber\\ \label{w7}
&& \makebox[.5cm]{}\mbox{sgn}(l-m)\mbox{sin}(q(l-m)) |lm,S\rangle ]
\\ \nonumber\\ \label{w7b}
\varepsilon_{kq} =&& 2[1+\lambda\cos(k/2)\cos(q)]
\end{eqnarray}
where $|lm,S\rangle=\sum_\alpha|t_{l\alpha}t_{m\alpha}\rangle/\sqrt{3}$
refers to the singlet combination of two triplets in real space and
$\varepsilon_{kq}$ is the bare two-particle kinetic energy. $k$($2q$) is
the total(relative) momentum, $\langle k'q'|kq\rangle
=\delta_{kk'}\delta_{qq'}$ and $k\in]-\pi,\pi]$, $q\in]0,\pi]$.
(Anti)periodic boundary conditions apply to $k$($q$) and $|k,q\rangle=-
|k,-q\rangle$. A discussion of the remaining two-particle states in the
triplet and quintet sector is deferred to appendix \ref{A}.

After these preliminaries we are in a position to evaluate the optical
absorption due to {\em phonon-assisted two-magnon emission} (PME).  This
absorption results from a magnetoelastic excitation induced by the incoming
photon-field and has been considered first in the planar geometry of the
CuO$_2$ square-lattice of the HT$_C$ materials
\cite{Lorenzana95a,Lorenzana95b}. Its microscopic justification can
be applied with little change to the case of spin-ladders with
corner-sharing Cu-O structure. For the effective coupling one obtains
\begin{eqnarray}
H_{LS} =
-E(t) \sum_{l,\alpha,\mu} [&& p_I u_{\mu Ol} + p_A (2 u_{\mu Ol}
\nonumber\\ \label{a1}
&& - u_{\mu Ll}-u_{\mu Rl})] S^\alpha_{\mu l}S^\alpha_{\mu l+1}
\end{eqnarray}
$E(t)$ is the time-dependent electric field which is polarized along the
ladder.  $u_{\mu O(L,R)l}$ denote displacement coordinates of the oxygen
site {\em on} the $\mu$-th leg of the ladder and refer to the oxygen sites
in cell $O(rigin)=l$, $L(eft)=l-1$, and $R(ight)=l+1$. The effective
charges $p_I$ and $p_A$ can be derived from an expansion up to first order
in $u_l$ and $E(t)$ of the Cu-O-Cu superexchange coupling. Expressions for
$p_I$ and $p_A$ can be found in \cite{Lorenzana95a,Lorenzana95b}. For
cuprates which are strongly covalent materials and for $E(t)$ polarized
parallel to the ladder, $|p_A|\gg |p_I|$ is a reasonable approximation
\cite{Lorenzana95b}.  Although the specific form of (\ref{a1}) depends on
the corner-sharing Cu-O structure it is conceivable that a similar PME
is also
possible in other spin-ladder compounds. Introducing the Fourier
transformed phonon-operators $u_{k\mu}=(a^{\phantom{\dagger}}_{k\mu} +
a^\dagger_{-k\mu}) / \sqrt{2M\omega_k}$ with the phonon dispersion
$\omega_k$ and the reduced O mass $M$ and replacing $S^\alpha_{\mu l}$ with
the bond variables $S^\alpha_l$ and $T^\alpha_l$ we get
\begin{eqnarray}
\label{a2}
H_{LS} =&& E(t) \frac{1}{\sqrt{2}}\sum_{k,\mu}
(a^{\phantom{\dagger}}_{k\mu}+a^\dagger_{-k\mu}) P_k
\\ \nonumber
P_k =&& - \frac{1}{4\sqrt{N M\omega_k}} \sum_{l,\alpha} e^{ikl}
[p_I + 2 p_A (1-\cos(k))] T^\alpha_l T^\alpha_{l+1}
\\ \label{a2b}
:= && - \frac{1}{\sqrt{N}} \sum_{l,\alpha} p_k e^{ikl}
T^\alpha_l T^\alpha_{l+1}
\end{eqnarray}
where terms proportional to $S^\alpha_lS^\alpha_{l+1}$ and
$S^\alpha_lT^\alpha_{l+1}$ have been dropped since they do not act on a
pure singlet product-state. Photon momentum conservation has been neglected
since the wave length of light is large compared to the lattice spacing.

The zero-temperature absorption coefficient $I(\omega)$ at frequency
$\omega$, normalized to the electric field intensity $|E|^2$, is obtained
from Fermi's golden rule. To simplify technical matters we assume only
optical phonons of frequency $\omega_k=\omega_0$ to be involved in the
absorption process. Since our model discards any intrinsic magnetoelastic
coupling the phonons represent a mere 'momentum bath' allowing for
arbitrary recoil of the two triplets \cite{TwoMagNote}. Tracing over the
phonons leads to
\begin{eqnarray}
\label{a3}
I(\omega)=&&2\pi \sum_{f,k} |\langle fk| P_k |0\rangle|^2
\delta(\omega-E_{fk}-\omega_0)
\\ \nonumber
=&& -2 \;\mbox{Im}\; \sum_{k,q,q'} [\langle 0| P^\dagger_k
|kq,S\rangle
\\  \label{a3b}
&& \makebox[.5cm]{}\langle kq,S|\frac{1}{z-H}
|kq',S\rangle\langle kq',S| P^{\phantom{\dagger}}_k|0\rangle]
\end{eqnarray}
where $z=\omega+i0^+-\omega_0$ and for the remainder of this work we redefine
$\omega-\omega_0\rightarrow\omega$. While $|fk\rangle$ are the {\em dressed}
two-particle eigenstates of the ladder of total momentum $k$, spin zero and
energy $E_{fk}$, $|kq,S\rangle$ are the bare states (\ref{w7}).

\begin{figure}[tb]
\vskip -4.5cm 
\centerline{\psfig{file=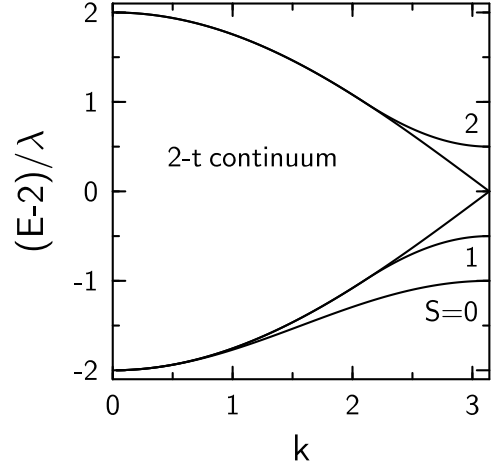,width=12cm}}
\vskip -4.3cm 
\caption[l]{Dispersion of the $S=0,1$, and $2$ bound states relative to the
bare two-triplet continuum.}
\label{fig1}
\end{figure}

Next we have to evaluate the resolvent (\ref{a3b}).  Since $|kq,S\rangle$
contains only two triplets and, as noted previously, we neglect the coupling
between sectors of different triplet number this task is equivalent to
solving an interacting two-particle problem which can be done exactly. To
this end we observe that
\begin{eqnarray}
\label{a4a}
\langle kq,S| H_3 |k'q',S\rangle &=& -\frac{4}{N-1}\delta_{kk'}\lambda
\sin(q)\sin(q')
\\ \label{a4b}
\langle k'q,S| P^{\phantom{\dagger}}_k |0\rangle
&=& -\sqrt{\frac{12}{N-1}} p_k e^{-ik/2} \sin(q) \delta_{kk'}
\end{eqnarray}
This can be combined with (\ref{a3b}) to sum the T-Matrix series for
$\langle kq,S|1/(z-H)|kq',S\rangle$ yielding
\begin{equation}
\label{a5}
I(\omega) = -24 \;\mbox{Im}\; \sum_{k}
\frac{p_k^2}{G^{-1}(k,\omega+i0^+)+4\lambda}
\end{equation}
where the bare two particle resolvent $G(k,z)$ is given by
\begin{eqnarray}
G(k,z)&=&\frac{1}{4\pi}\int^\pi_{-\pi}dq
\frac{\sin^2(q)}{z-(2+2\lambda\cos(k/2)\cos(q))}
\nonumber\\[.5cm]\label{a6}
&=& \frac{\mbox{sign}[\mbox{Re}(a)]\sqrt{a^2-1}-a}{4\lambda\cos(k/2)}
\end{eqnarray}
with $a=(2-z)/[2\lambda\cos(k/2)]$.
The zeros at $\omega=E^S_k$ and momentum $k$ of the denominator on the
r.h.s. of (\ref{a5})
\begin{equation}
\label{a7}
1+4\lambda G(k,E^S_k)=0
\end{equation}
correspond to (anti)bound states in the two-triplet spin-singlet sector.
Since both, the boundaries of the two-particle continuum (\ref{w7b}) as well
as the solutions of (\ref{a7}) are functions of $k$ only if energies
$\omega$ are rescaled in terms of the quantity $(\omega-2)/\lambda$, the
$k$-space structure of the rescaled two-particle spectrum is {\em
independent} of $\lambda$ within our approximation. In particular, a singlet
bound-state exists at all $k$ in the Brillouin zone. At $k=\pi$ its binding
energy is largest with $E^S_k=2-\lambda$ while at $k=0$ the binding energy
is zero.  This is shown in fig. \ref{fig1}, which depicts the bound-state
dispersion along with the two-particle continuum. Also shown are triplet
($S=1$) and quintet ($S=2$) (anti)bound-states, which are optically
inactive. These states are discussed in appendix \ref{A}. Recently,
similar collective states have been predicted using bond-boson techniques
\cite{Damle98,Kotov98,UinftyNote}.

\begin{figure}[t]
\vskip -4.5cm 
\centerline{\psfig{file=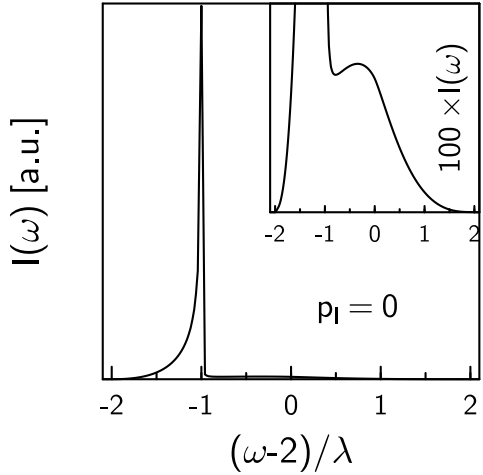,width=12cm}}
\vskip -4.1cm 
\caption[l]{Optical absorption spectrum for $p_I=0$ .
Inset: identical spectrum, y-axis stretched by 100.}
\label{fig2}
\end{figure}

\begin{figure}[t]
\vskip -4.5cm 
\centerline{\psfig{file=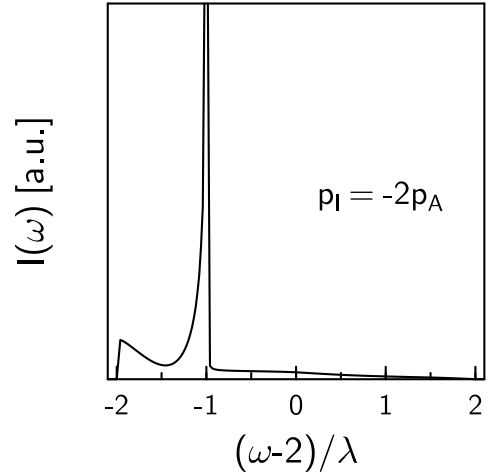,width=12cm}}
\vskip -4.2cm 
\caption[l]{Optical absorption spectrum for $p_I=-2 p_A$.}
\label{fig3}
\end{figure}

From (\ref{a4b},\ref{a5},\ref{a6}) we can obtain the absorption by
numerical integration. The rescaled intensity $\lambda
I(\tilde{\omega})/p_A$ is a function of $\tilde{\omega}=(\omega-2)/\lambda$
and $p_I/p_A$ only and does not dependent on $\lambda$. Figure \ref{fig2}
depicts the absorption spectrum for $p_I=0$. We note that our choice of the
frequency variable $\tilde{\omega}$ implies that zero incoming photon
energy corresponds to the point $-2/\lambda$ on the x-axis of
fig. \ref{fig2}. For any finite spin gap, i.e. $\lambda<1$, this point is
off the range plotted which refers only to the frequency window
$-2\leq\tilde{\omega}\leq 2$ in which absorption occurs. Figure \ref{fig2}
demonstrates that the bound state has a profound impact on the absorption
spectrum which comprises almost completely of a structure due to the
integrated density of states of the bound state. In particular there is a
van-Hove singularity at $\omega=2-\lambda$ which results from the
maximum in the bound-state dispersion at $k=\pi$. Only in the special case
of $p_I=-4 p_A$ this van-Hove singularity disappears. The inset in
fig. \ref{fig2} focuses on the remaining spectral weight. It has a maximum
at the center of the two-triplet continuum, however its weight is very
small as compared to the contribution from the bound states. At finite
values of $p_I$ a step-like feature appears at the spin-gap and for $p_I<0$
a dip occurs due to interference in the effective charge $p_k$. At present
we believe that such features are of academic interest only. Yet, for
completeness a typical spectrum displaying the latter effects is shown in
fig. \ref{fig3}.

The preceding discussion does not rule out an interpretation of van-Hove
like structures in an experimentally determined absorption spectrum solely in
terms the momentum integrated {\em bare} two-triplet continuum rather than in
terms of bound states. However, taking this point of view and using that
$|p_I/p_A|\ll 1$, which is most likely the case, the absorption spectrum
would rather be symmetric with respect to the center of gravity of the
measured spectrum. Finally we note, that in case of a finite dispersion of
the phonon additional structures may appear which are not contained in our
present analysis.

In conclusion we propose that singlet bound-states have a dramatic
impact on the optical absorption of dimerized spin-ladder systems which
allow for phonon-assisted multi-magnon absorption of the Lorenzana-Sawatzky
type. We hope that our work may stimulate further experimental investigation
of this issue.

One of us (W.B.) acknowledges the kind hospitality of the ITP at UC Santa
Barbara. This research was supported in part by the Deutsche
Forschungsgemeinschaft under Grant No. BR 1084/1-1 and the National Science
Foundation under Grant No. PHY94-07194.

\begin{appendix}
\section{S=1,2 states}\label{A}
The triplet and quintet two-particle states in real space correspond to the
combinations $|lm,\alpha\rangle=$ $\sum_{\beta,\gamma}$
$\varepsilon_{\alpha\beta\gamma}$ $|t_{l\beta} t_{m\gamma}\rangle/\sqrt{2}$
and $|lm,\alpha
\beta\rangle =\sqrt{3/8}(|t_{l\alpha} t_{m\beta}\rangle+|t_{l\beta}
t_{m\alpha}\rangle-\delta_{\alpha\beta} (2/3) \sum_{\gamma}|t_{l\gamma}
t_{m\gamma}\rangle)$ respectively. Due to spin rotational invariance we may
focus on the total-$S^z=0$ components only, i.e. $|lm,P\rangle=|lm,
z\rangle$ (triplet) and $|lm,D\rangle=|lm,zz\rangle$ (quintet). The bare
two-particle eigenstates are
\begin{eqnarray}
\label{appa1}
|kq,P\rangle = && \frac{1}{N} \sum_{l,m} e^{i[k(l+m)/2 +q(l-m)]}
|lm,P\rangle
\\
|kq,D\rangle = && \frac{1}{\sqrt{N(N-1)}} \sum_{l,m} [
e^{ik(l+m)/2}
\nonumber\\ \label{appa1b}
&&  \makebox[.5cm]{}\mbox{sgn}(l-m)\mbox{sin}(q(l-m)) |lm,D\rangle ]
\end{eqnarray}
where identical properties regarding the momenta as in (\ref{w7}) apply and
the bare two-particle energies of the states $|kq,P\rangle$ and
$|kq,D\rangle$ are identical to $\varepsilon_{kq}$ of (\ref{w7}). Note that
both, $P$- and $D$-states conform with the hardcore constraint
$|t_{l\alpha} t_{l\beta}\rangle=0$. From the identity $({\bf S}_l+{\bf
S}_m)^2/2-2={\bf S}_l\cdot{\bf S}_m$ and from (\ref{a4a}) it follows
directly that
\begin{equation}
\label{appa2}
\langle kq,X| H_3 |k'q',X\rangle =
-\frac{c_X}{N-1} \delta_{kk'}\lambda\sin(q)\sin(q')
\end{equation}
with $c_X=4$, $2$, and $-2$ for $X=S$, $P$, and $D$ respectively.
Therefore the energies $E^X_k$ of the (anti)bound states in all spin
channels are obtained from the zeros of the single equation
\begin{equation}
\label{appa3}
1+c_X\lambda G(k,E^X_k)=0
\end{equation}
with $G(k,z)$ as in (\ref{a6}) and $c_X$ set according to the total spin.
For $X=P(D)$ a(n) (anti)bound state exists for $k\geq 2\pi/3$ for all
values of $\lambda$ with $E^{P(D)}_{\pi}=2-(+)\lambda/2$. In
fig. \ref{fig1} the dispersion of these (anti)bound states relative to the
two-triplet continuum is depicted.
\end{appendix}


\begin{thebibliography}{99}
\bibitem{TwoMagNote}
This is different from two-magnon Raman scattering which probes
zero total two-magnon momentum only and may therefore be
less well suited to observe $S=0$ bound states with dominant
weight at the zone boundary.
\bibitem{UinftyNote}
While in the present work two triplets never occupy the same site by
construction of (\ref{w7},\ref{appa1},\ref{appa1b}) a ficiticious hardcore
potential has to be dealt with in refs. \cite{Damle98,Kotov98} in order to
remove unphysical states containing double-occupancies from the Hilbert space.
\end{thebibliography}
\end{document}